\documentclass{ws-mpla}

\usepackage[super]{cite}
\usepackage{xcolor}
\usepackage[verbose,hypertexnames=false]{hyperref}
\usepackage{physics}
\hypersetup{colorlinks=false,allbordercolors=blue,pdfborderstyle={/S/U/W 1}}

\usepackage[normalem]{ulem}  
\usepackage{color} 

\renewcommand\sout{\bgroup \color{red} \ULdepth=-.5ex \ULset}

\begin{document}

\markboth{K.~Sone,  T.~R.~Hu, KF.~K.~Guo, T.~Hyodo and I.~Low}{Entanglement suppression for $\Omega\Omega$ scattering}

\catchline{}{}{}{}{}

\title{Entanglement suppression for $\Omega\Omega$ scattering
}

\author{Katsuyoshi Sone}
\address{
Department of Physics, Tokyo Metropolitan University,\\
Hachioji 192-0397, Japan\\
sone-katsuyoshi@ed.tmu.ac.jp}

\author{Tao-Ran Hu}
\address{
School of Physical Sciences, University of Chinese Academy of Sciences,\\
Beijing 100049, China\\
hutaoran21@mails.ucas.ac.cn
}

\author{Feng-Kun Guo}
\address{
School of Physical Sciences, University of Chinese Academy of Sciences,\\
Beijing 100049, China\\
Institute of Theoretical Physics, Chinese Academy of Sciences,
\\ Beijing 100190, China\\
fkguo@itp.ac.cn
}

\author{Tetsuo Hyodo}
\address{
Research Center for Nuclear Physics (RCNP), The University of Osaka,\\
Ibaraki, Osaka 567-0047, Japan\\
hyodo@rcnp.osaka-u.ac.jp
}

\author{Ian Low}
\address{
High Energy Physics Division, Argonne National Laboratory,\\
Argonne, IL 60439, USA\\
Department of Physics and Astronomy, Northwestern University,\\
Evanston, IL 60208, USA\\
ilow@northwestern.edu
}

\maketitle

\begin{history}
\received{(Day Month Year)}
\revised{(Day Month Year)}
\accepted{(Day Month Year)}
\published{(Day Month Year)}
\end{history}

\begin{abstract}
We study entanglement suppression in $s$-wave $\Omega\Omega$ scattering, where each baryon has spin $3/2$. By treating the $S$-matrix as a quantum operator acting on the spin states, we quantify its ability to generate entanglement and identify the conditions on the phase shifts of the spin channels that minimize entanglement generation in the system. In $\Omega\Omega$ scattering, only antisymmetric spin channels are allowed due to Fermi-Dirac statistics. Applying the entanglement-suppression framework to $\Omega\Omega$ scattering, we find two solutions for the phase shifts: one leading to a spin SU(4) symmetry and the other to a nonrelativistic conformal symmetry. We show that the solution associated with the nonrelativistic conformal symmetry originates from the specific structure of the Clebsch-Gordan coefficients in the $3/2 \otimes 3/2$ system.
\end{abstract}


\ccode{PACS Nos.: 03.65.Nk, 03.65.Mn, 13.75.Cs}

\section{Introduction}

Symmetry plays a central role in understanding hadronic interactions. A prominent example is nucleon-nucleon ($NN$) scattering, which is constrained by the spin-SU(2) and isospin-SU(2) symmetries of QCD. Within the ${\rm SU(2)} \times {\rm SU(2)}$ framework, low-energy $NN$ scattering can be parameterized by two independent components. Phenomenologically, however, these two components are known to exhibit very similar properties, indicating an enhanced spin-flavor SU(4) symmetry\cite{Wigner:1936dx}. Moreover, the large scattering lengths in these channels imply that the system exhibits the nonrelativistic conformal invariance~\cite{Mehen:1999nd}. 

One explanation for the emergence of the spin-flavor SU(4) symmetry has been given in the context of large-$N_c$ QCD~\cite{Kaplan:1995yg}. Recently, an alternative approach based on entanglement suppression has been proposed~\cite{Beane:2018oxh} as a conjectural principle for identifying emergent symmetries directly from scattering processes. By regarding the $S$-matrix as a quantum operator acting on a two-baryon state, its ability to generate entanglement in the system can be characterized through the entanglement power~\cite{Zanardi:2001zza}. The minimization of the entanglement power selects special quantum logic gates, notably the identity and SWAP gates~\cite{Low:2021ufv}. In the case of $NN$ scattering, these minimal-entanglement solutions are found to be associated with the enhanced symmetry structures~\cite{Beane:2018oxh, Low:2021ufv}: the identity gate is associated with the spin-flavor SU(4) symmetry, while the SWAP gate corresponds to nonrelativistic conformal symmetry. The idea of entanglement suppression has been applied to various hadron scattering systems~\cite{Beane:2018oxh, Low:2021ufv, Liu:2022grf, Liu:2023bnr, Kirchner:2023dvg, Hu:2024hex}.

In this work, we apply the entanglement suppression framework to $\Omega\Omega$ scattering~\cite{Hu:2025lua}, in which each baryon carries spin $3/2$. The $\Omega$ baryon is the only member of the decuplet that is stable under the strong interaction, which allows one to realize a well-defined scattering of spin-$3/2$ particles in a hadronic system. In addition, lattice QCD studies have reported that the total spin $J=2$ channel of the $\Omega\Omega$ system possesses a weakly bound state near the threshold, resulting in a large scattering length~\cite{Gongyo:2017fjb}. We study how entanglement suppression constrains the structure of the $\Omega\Omega$ $S$-matrix and discuss the emergent symmetries that arise from this constraint in detail.

\section{Entanglement suppression for $NN$ scattering}
\label{sec: formulation}

We introduce entanglement suppression as a conjecture to find emergent symmetries in hadron-hadron scattering. In this section, we review the application of this framework to $NN$ scattering~\cite{Beane:2018oxh, Low:2021ufv} and examine the associated emergent symmetries.

\subsection{$S$-matrix}

Here we consider the $S$-matrix for low-energy $s$-wave $NN$ scattering in order to discuss entanglement suppression. A nucleon has two spin states: spin-up $\ket{\uparrow}$ or spin-down $\ket{\downarrow}$. For a two-nucleon system, the total spin is given by the decomposition
\begin{align}
    \frac{1}{2} \otimes \frac{1}{2}
    =
    0 \oplus 1,
    \label{eq: decom spin}
\end{align}
where $J=1$ corresponds to a symmetric irreducible representation (irrep), whereas $J=0$ corresponds to an antisymmetric one. In addition to the spin degree of freedom, a nucleon possesses two isospin states, $\ket{p}$ and $\ket{n}$, forming an isospin doublet. The total isospin can also be decomposed as in Eq.~\eqref{eq: decom spin}, yielding two isospin states $I=0$ and $I=1$. Due to Fermi-Dirac (FD) statistics, only two spin-isospin channels $(J,I)=(0,1)$ and $(1,0)$ are allowed in $NN$ scattering.

Based on this observation, the $S$-matrix for low-energy $s$-wave $NN$ scattering is given by
\begin{align}
    \hat{S}_{NN}
    =
    e^{2i\delta_{0}} \mathcal{J}_0
    +
    e^{2i\delta_{1}} \mathcal{J}_1,
    \label{eq: Smat NN}
\end{align}
where $\delta_0$ and $\delta_1$ are the phase shifts in the spin channels $J=0$ and $J=1$, respectively. The operators $\mathcal{J}_0$ and $\mathcal{J}_1$ denote the corresponding projection operators which can be expressed in terms of the Pauli matrices $\bm{\sigma}$, which are the generators of SU(2):
\begin{align}
    \mathcal{J}_0 
    =
    \frac{1}{4}\left(1 - \bm{\sigma}\cdot\bm{\sigma}\right),
    &\quad
    \mathcal{J}_1
    =
    \frac{1}{4}\left(3 + \bm{\sigma}\cdot\bm{\sigma}\right),
    \quad \bm{\sigma}\cdot\bm{\sigma}
    =
    \sum_{i=1}^{3} \sigma_{i} \otimes \sigma_{i} .
\end{align}
In Eq.~\eqref{eq: Smat NN}, only the spin projectors are included, because the total isospin is uniquely determined for a given spin state due to FD statistics.

\subsection{Entanglement measure}

Within the framework of entanglement suppression, the quantum correlations between two hadrons are quantified by an entanglement measure. In this work, we employ the linear entropy to quantify entanglement for a given state.
The linear entropy for a two-particle state $\ket{\psi} \in \mathcal{H}_{12}$, where the Hilbert space $\mathcal{H}_{12}$ is given by the tensor product $\mathcal{H}_{12} = \mathcal{H}_1 \otimes \mathcal{H}_2$, is defined as
\begin{align}
    \mathcal{E}(\ket{\psi})
    &=
    \Tr_{1}\rho_1 - \Tr_{1}\rho_1^2,
    \label{eq: def of ent}
\end{align}
where $\rho_1$ is the reduced density matrix obtained by tracing out subsystem 2:
\begin{align}
    \rho_1 \equiv \Tr_2 \rho, \qquad
    \rho = \ket{\psi}\bra{\psi}.
\end{align}

For a bipartite system consisting of two spin-$J$ particles, a normalized state can be written in general as
\begin{align}
    \ket{\psi}
    &=
    \sum_{i,j}
    \alpha_{ij} \ket{ij}, \quad 
    \ket{ij}=\ket{i}\otimes\ket{j},\quad
    \sum_{i,j} |\alpha_{ij}|^2 = 1,
    \label{eq: gne state}
\end{align}
where $i$ and $j$ run from $1$ to $2J+1$ and specify the third component of the spin $J_3$ as
\begin{align}
    \ket{1}=\ket{J_3=J},\quad 
    \ket{2}=\ket{J_3=J-1},\quad 
    \cdots,\quad 
    \ket{2J+1}=\ket{J_3=-J}.
\end{align}
The coefficient $\alpha_{ij}$ is given by $\alpha_{ij}=\bra{ij}\ket{\psi}$.
Using the definition in Eq.~\eqref{eq: def of ent}, the linear entropy of the state $\ket{\psi}$ in Eq.~\eqref{eq: gne state} can be expressed in terms of the coefficients $\alpha_{ij}$ as
\begin{align}
    \mathcal{E}(\ket{\psi})
    =
    2\sum_{i<i^{\prime},\,j<j^{\prime}}
    \qty|\alpha_{ij}\alpha_{i^{\prime}j^{\prime}}-\alpha_{i^{\prime}j}\alpha_{ij^{\prime}}|^2.
    \label{eq: entropy general}
\end{align}

\subsection{Entanglement power}

The entanglement of a two-particle system may increase through the scattering process. In other words, the $S$-matrix can generate entanglement between the two particles.
To quantify the ability of the $S$-matrix to generate entanglement, we introduce the entanglement power~\cite{Zanardi:2001zza} (EP), defined as
\begin{align}
    E(\hat{S})
    &=
    \overline{\mathcal{E}(\hat{S}\ket{\psi_{\rm in}})}
    =
    \int d\omega_1 d\omega_2\; \mathcal{E}(\hat{S}\ket{\psi_{\rm in}}),
    \label{eq: def of EP}
\end{align}
where the overline denotes the average over the initial states.
In the definition of the EP in Eq.~\eqref{eq: def of EP}, the initial state $\ket{\psi_{\rm in}}$ is taken to be a product state of two spin states,
\begin{align}
    \ket{\psi_{\rm in}}
    =
    \ket{\psi_1}\otimes\ket{\psi_2}
    =
    \qty(\sum_{i}u_i\ket{i})
    \otimes
    \qty(\sum_{j}v_j\ket{j}),
    \label{eq: def of in}
\end{align}
where the integration measure $d\omega_{1,2}$ corresponds to the uniform average over single-particle spin states.
By comparing Eqs.~\eqref{eq: gne state} and~\eqref{eq: def of in}, one finds that the coefficients $\alpha_{ij}$ for the initial state $\ket{\psi_{\rm in}}$ are separable:
\begin{align}
    \bra{ij}\ket{\psi_{\rm in}}
    &=
    \alpha_{ij}
    =
    u_i v_j.
    \label{eq: sepaalpha}
\end{align}
Substituting Eq.~\eqref{eq: sepaalpha} into Eq.~\eqref{eq: entropy general}, it can be verified that the linear entropy of the product state vanishes.

In general, a single-particle spin state $\sum_i u_i \ket{i}$ in a $(2J+1)$-dimensional Hilbert space is represented by a point on the complex projective manifold $\mathbb{CP}^{m}$ with $m=2J$. The $(m+1)$-component complex vector $\bm{u}$ can be parameterized by $2m$ angles $(\theta_i,\nu_i)$ with $i=1,\dots,m$ as~\cite{Bengtsson:2001yd,Bengtsson}:
\begin{align}
    \bm{u}
    &=
    \begin{pmatrix}
        n_0, & n_1 e^{i\nu_1}, \dots, & n_m e^{i\nu_m}
    \end{pmatrix},
    \quad 
    0 \le \nu_i \le 2\pi, \\
&\left. \left\{ \begin{matrix}
	n_0 &=& \sin \theta_1 \sin \theta_2 \sin \theta_3 \cdots \sin \theta_m, \\
	n_1 &=& \cos \theta_1 \sin \theta_2 \sin \theta_3 \cdots \sin \theta_m, \\
	n_2 &=& \cos \theta_2 \sin \theta_3 \cdots \sin \theta_m, \\
	\vdots && \vdots \\
	n_m &=& \cos \theta_m,
\end{matrix} \right. \right.,
\quad
0 \le \theta_i \le \frac{\pi}{2}.
\end{align}
In Eq.~\eqref{eq: def of EP}, the integral is performed over all directions of each single-particle spin state using
\begin{equation}
\mathrm{d}\omega
=
\frac{m!}{\pi^m}
\prod_{i=1}^{m}
\int_{0}^{\pi/2} \mathrm{d}\theta_i
\int_{0}^{2\pi} \mathrm{d}\nu_i\,
\cos\theta_i\,\sin^{2i-1}\theta_i,
\end{equation}
which is known as the Fubini--Study measure~\cite{Bengtsson:2001yd,Bengtsson}.

Using the $S$-matrix~\eqref{eq: Smat NN} and the definition of the EP in Eq.~\eqref{eq: def of EP}, we obtain the EP for $NN$ scattering as
\begin{align}
    E(\hat{S}_{NN})
    &=
    \frac{1}{6}\sin^2\!\left(2\delta_{01}\right),\quad
    \delta_{01}\equiv \delta_{0}-\delta_{1}.
    \label{eq: EP NN}
\end{align}
Because the phase shifts are real, we find that the EP is nonnegative, $E(\hat{S}_{NN})\geq 0$.

\subsection{Emergent symmetries}
\label{eq: NN scattering}

In this section, we discuss the emergent symmetries corresponding to solutions minimizing the EP. The minimal value of $E(\hat{S}_{NN})$ in Eq.~\eqref{eq: EP NN} is zero, which is attained when
\begin{align}
    |\delta_{01}|=0,\ \frac{\pi}{2}.
    \label{eq: solu NN}
\end{align}
For the solution $|\delta_{01}|=0$, the $S$-matrix of $NN$ scattering reads
\begin{align}
    \hat{S}_{NN} \propto \mathcal{J}_0 + \mathcal{J}_1 = \hat{1}.\label{eq:NNidentity}
\end{align}
This $S$-matrix corresponds to the identity gate in quantum information theory.
For $\hat{S}_{NN}\propto\hat{1}$, the final state is identical to the initial state up to an overall phase: $\ket{\psi_{\rm out}}\equiv\hat{S}_{NN}\ket{\psi_{\rm in}}\propto\ket{\psi_{\rm in}}$. The condition $|\delta_{01}|=0$ corresponds to $\delta_{0}=\delta_{1}$, indicating that the interactions in the spin channels $J=0$ and $J=1$ are identical. Therefore, this solution realizes the spin-flavor SU(4) symmetry.

The solution $|\delta_{01}|=\pi/2$ gives the $S$-matrix
\begin{align}
    \hat{S}_{NN}\propto\mathcal{J}_0-\mathcal{J}_1
    =\frac{1}{2}\left(1 + \bm{\sigma}\cdot\bm{\sigma}\right)
    \equiv \operatorname{SWAP},
    \label{eq: SWAP NN}
\end{align}
which is referred to as the SWAP gate. The SWAP operator exchanges the two spin states as
\begin{align}
    \operatorname{SWAP}\ket{\psi_1}\otimes\ket{\psi_2}
    =
    \ket{\psi_2}\otimes\ket{\psi_1}.
\end{align}
The condition $|\delta_{01}|=\pi/2$ at zero momentum indicates that one of the two spin channels reaches the unitarity limit, while the other exhibits trivial scattering with no interaction. At the unitarity limit, the scattering length diverges, whereas it vanishes in the case of non-interacting scattering. The absence of a length scale in low-energy scattering induces the nonrelativistic conformal symmetry~\cite{Mehen:1999nd}. This symmetry is therefore associated with the $S$-matrix exhibiting the SWAP structure.

Both the identity and the SWAP gates map a product state onto a product state. As a result, the EP of the $S$-matrix vanishes, since the final state $\ket{\psi_{\rm out}}=\hat{S}_{NN}\ket{\psi_{\rm in}}$ remains separable for any initial product state. It can be shown that the identity and the SWAP gates are the only operators that yield the minimal entanglement power~\cite{Low:2021ufv}.

\section{Application to $\Omega\Omega$ scattering}
\label{sec: Omega application}

In this section, we discuss entanglement suppression in $\Omega\Omega$ scattering and the emergent symmetries of the system. In $\Omega\Omega$ scattering, there are two spin channels, $J=0$ and $J=2$. Lattice QCD calculations report that the scattering length of the $J=2$ channel is $a_{J=2}\approx 4.6\ {\rm fm}$~\cite{Gongyo:2017fjb}, which is much larger than the typical length scale of the strong interaction, about 1~fm. Motivated by this observation, we regard the $J=2$ channel as being close to the unitarity limit, $\delta_{J=2}\approx\pi/2$, and analyze the behavior of the $J=0$ channel based on entanglement suppression.

\subsection{$S$-matrix}

Here we introduce the $S$-matrix for the $s$-wave $\Omega\Omega$ scattering. Since the $\Omega$ baryon carries spin $3/2$, the total spin of the $\Omega\Omega$ system is classified as
\begin{align}
    \frac{3}{2}\otimes\frac{3}{2}
    =
    0\oplus1\oplus2\oplus3.
\end{align}
Among these, $J=0$ and $J=2$ correspond to antisymmetric irreducible representations, whereas $J=1$ and $J=3$ are symmetric ones. In the $s$-wave scattering of identical fermions such as the $\Omega\Omega$ system, the spin wave function must be antisymmetric due to FD statistics.
Therefore, only the $J=0$ and $J=2$ channels are allowed in $\Omega\Omega$ scattering, while the symmetric states with $J=1$ and $J=3$ are forbidden in the low-energy limit.

The low-energy $S$-matrix for $\Omega\Omega$ scattering is therefore given by
\begin{align}
    \hat{S}_{\Omega\Omega}
    =
    e^{2i\delta_{0}} \mathcal{J}_0
    +
    e^{2i\delta_{2}}\mathcal{J}_2,
    \label{eq: Smat for omega}
\end{align}
where $\delta_{0}$ and $\delta_{2}$ are the phase shifts in the $J=0$ and $J=2$ channels, respectively. The projection operators $\mathcal{J}_{J}$ for $J=0,1,2,3$ can be expressed as polynomials of the Casimir operator $\boldsymbol{t}_{3/2}\cdot\boldsymbol{t}_{3/2}$:
\begin{align}
   \mathcal{J}_J
   =
   \prod_{J^\prime \neq J}
   \frac{\boldsymbol{t}_{3/2}\cdot\boldsymbol{t}_{3/2}-\lambda_{J^\prime}}
        {\lambda_J-\lambda_{J^\prime}} .
\end{align}
Here, $\boldsymbol{t}_{3/2}$ denotes the SU(2) generators in the four-dimensional representation, and $\lambda_J$ denotes the eigenvalue of $\boldsymbol{t}_{3/2}\cdot\boldsymbol{t}_{3/2}$ for the state with total spin $J$. Explicit expressions of $\mathcal{J}_J$ can be found in Ref.~\citen{Hu:2025lua}.

\subsection{Entanglement power}
\label{sec: EP dd oo}

In this section, we consider the minimal EP solutions for $\Omega\Omega$ scattering. However, the EP formula~\eqref{eq: def of EP} cannot be directly applied to $\Omega\Omega$ scattering due to the non-unit normalization of the final state~\cite{Hu:2025lua}. Because the spin channels $J=1,3$ are absent, the final state obtained from an initial product state~\eqref{eq: def of in},
$\hat{S}_{\Omega\Omega}\ket{\psi_{\rm in}}$, is not normalized:
\begin{align}
    \bra{\psi_{\rm in}}\hat{S}_{\Omega\Omega}^{\dagger}\hat{S}_{\Omega\Omega}\ket{\psi_{\rm in}}
    &=
    \bra{\psi_{\rm in}}\mathcal{A}\ket{\psi_{\rm in}}
    =
    1 - \bra{\psi_{\rm in}}\mathcal{S}\ket{\psi_{\rm in}}
    <
    1,
    \\
    \mathcal{A}
    &\equiv
    \mathcal{J}_0+\mathcal{J}_2,
    \quad 
    \mathcal{S}
    \equiv
    \mathcal{J}_1+\mathcal{J}_3,
    \label{eq: def of A}
\end{align}
where $\mathcal{A}$ ($\mathcal{S}$) denotes the sum of the projection operators for all antisymmetric (symmetric) spin channels in the system. We therefore introduce the normalized final state
\begin{align}
    \ket{\psi^{\Omega\Omega}_{\rm out}}
    =
    \frac{\hat{S}_{\Omega\Omega}\ket{\psi_{\rm in}}}
    {\sqrt{\bra{\psi_{\rm in}}\mathcal{A}\ket{\psi_{\rm in}}}},
    \label{eq: norm final}
\end{align}
which satisfies
$\braket{\psi^{\Omega\Omega}_{\rm out}}=1$.

With this normalized state, the EP for $\Omega\Omega$ scattering is defined as
\begin{align}
    E(\hat{S}_{\Omega\Omega})
    &=
    \overline{\mathcal{E}(\ket{\psi^{\Omega\Omega}_{\rm out}})}
    =
    1-\int d\omega_1 d\omega_2\;
    \frac{\Tr_{1}\rho_{1}^{2}}
    {\bra{\psi_{\rm in}}\mathcal{A}\ket{\psi_{\rm in}}^{2}},
    \label{eq:EPOmega}
\end{align}
where $\rho_{1}=\Tr_{2}[\hat{S}_{\rm \Omega\Omega}\ket{\psi_{\rm in}}\bra{\psi_{\rm in}}\hat{S}_{\rm \Omega\Omega}^{\dag}]$. Using Eq.~\eqref{eq: def of in} and noting the action of the antisymmetrization operator $\mathcal{A}\ket{\psi_{\rm in}}=\sum_{i,j}(u_{i}v_{j}-u_{j}v_{i})/2\ket{ij}$, the normalization factor $\bra{\psi_{\rm in}}\mathcal{A}\ket{\psi_{\rm in}}$ can be expressed in terms of $u_i$ and $v_j$ as
\begin{align}
    \bra{\psi_{\rm in}}\mathcal{A}\ket{\psi_{\rm in}}
    &=
    \sum_{i<i^{\prime}}
    \frac{|u_i v_{i^{\prime}} - u_{i^{\prime}} v_i|^2}{2}.
    \label{eq: norm A}
\end{align}
Because this factor depends on the spin directions of $\ket{\psi_{\rm in}}$ and appears in the denominator of Eq.~\eqref{eq:EPOmega}, it causes complications in the angular integration. It is therefore convenient to introduce the $k$-th weighted EP and choose $k=2$~\cite{Hu:2025lua}, which leads to
\begin{align}
    E_{2}(\hat{S}_{\Omega\Omega})
    &=
    1
    -
    \frac{\int d\omega_1 d\omega_2\; \Tr_{1}\rho_{1}^{2}}
    {\int d\omega_1 d\omega_2\;
    \bra{\psi_{\rm in}}\mathcal{A}\ket{\psi_{\rm in}}^{2}}.
    \label{eq:EP2Omega}
\end{align}
Substituting the $S$-matrix in Eq.~\eqref{eq: Smat for omega} into this expression, the weighted EP for $\Omega\Omega$ scattering is obtained as~\cite{Hu:2025lua}
\begin{align}
   E_{2}(\hat{S}_{\Omega\Omega})
   &=
   \frac{1}{48}
   \left(
   25 - \cos[4\delta_{02}]
   \right),
   \quad
   \delta_{02}
   \equiv
   \delta_{0} - \delta_{2}.
   \label{eq: EP for omega}
\end{align}
Equation~\eqref{eq: EP for omega} shows that the minimal value of $E_{2}(\hat{S}_{\Omega\Omega})$ is $1/2$, which is achieved when
\begin{align}
    |\delta_{02}|
    =
    0,\ \frac{\pi}{2}.
\end{align}
Unlike the EP for $NN$ scattering~\eqref{eq: EP NN}, the minimal value of the EP for $\Omega\Omega$ scattering is not zero, reflecting the absence of the symmetric spin channels $J=1$ and $J=3$.

\subsection{Emergent symmetries}
\label{sec:ESOmegaOmega}

Next, we discuss the emergent symmetries of $\Omega\Omega$ scattering. For $|\delta_{02}|=0$, the $S$-matrix reads
\begin{align}
    \hat{S}_{\Omega\Omega}
    \propto
    \mathcal{J}_0+\mathcal{J}_2
    =
    \mathcal{A}.
\end{align}
Noting that $\mathcal{A}$ acts as an identity operator within the subspace of antisymmetric states, we find this result analogous to Eq.~\eqref{eq:NNidentity} in NN scattering. Under the condition $|\delta_{02}|=0$, the system exhibits a spin SU(4) symmetry, in which the four spin components $J_3=\pm 3/2, \pm 1/2$ of a single $\Omega$ baryon furnish the fundamental representation of SU(4). The $J=0$ and $J=2$ combinations of two spins then furnish the antisymmetric irrep $\boldsymbol{6}$ of spin SU(4). Given the lattice QCD indication that the spin $J=2$ channel is close to the unitary limit, $\delta_{2}\approx\pi/2$, the condition $|\delta_{02}|=0$ implies that the $J=0$ channel is also close to the unitary limit, $\delta_{0}\approx\pi/2$.

For the solution $|\delta_{02}|=\pi/2$, the $S$-matrix reads
\begin{align}
    \hat{S}_{\Omega\Omega}
    \propto
    \mathcal{J}_2-\mathcal{J}_0
    \equiv
    \operatorname{SWAP}_{\mathrm{A}}.
\end{align}
We note that the operator $\operatorname{SWAP}_{\mathrm{A}}$ is distinct from the SWAP gate in $NN$ scattering~\eqref{eq: SWAP NN}, since $\operatorname{SWAP}_{\mathrm{A}}$ does not exchange the spin states of the two particles. The detailed properties and interpretation of the operator $\operatorname{SWAP}_{\mathrm{A}}$ will be discussed in the next section. By an argument analogous to that in Sec.~\ref{eq: NN scattering}, imposing the condition $|\delta_{02}|=\pi/2$ in the low-momentum limit implies that either the $J=0$ or $J=2$ channel reaches the unitary limit, where the corresponding scattering length diverges. Namely, this solution realizes an emergent nonrelativistic conformal symmetry. Assuming $\delta_{2}\approx \pi/2$ as suggested by lattice QCD, this solution implies that the $J=0$ channel is nearly non-interacting, $\delta_{0}\approx 0$.

It is instructive to compare this result with the EP for deuteron-deuteron ($dd$) scattering.
Because the deuteron $d$ is a spin-1 boson, $dd$ scattering involves only the spin channels $J=0$ and $J=2$ due to Bose-Einstein statistics.
In this case, however, the $J=0$ and $J=2$ channels correspond to symmetric irreps,
whereas these channels are antisymmetric in $\Omega\Omega$ scattering.
The EP for $dd$ scattering is calculated as~\cite{Kirchner:2023dvg}
\begin{align}
    E_{2}(\hat{S})
    &= 
    \frac{1}{594}\qty(171-70\cos[2\delta^{dd}_{02}] - 20\cos[4\delta^{dd}_{02}]),
    \quad \delta^{dd}_{02} \equiv \delta^{dd}_0-\delta^{dd}_2,
    \label{eq: EF for dd}
\end{align}
where $\delta^{dd}_{0}$ and $\delta^{dd}_{2}$ are the phase shifts in the spin $J=0$ and $J=2$ channels of $dd$ scattering, respectively.
Equation~\eqref{eq: EF for dd} shows that the solution $|\delta^{dd}_{02}|=\pi/2$ yields a larger EP than $|\delta^{dd}_{02}|=0$,
due to the presence of the $\cos[2\delta^{dd}_{02}]$ term,
in contrast to the case of $\Omega\Omega$ scattering.
In Sec.~\ref{eq: sp solu}, we discuss the reason why the solution $|\delta_{02}|=\pi/2$ gives the same EP with $|\delta_{02}|=0$ in $\Omega\Omega$ scattering.

\subsection{Property of ${\rm SWAP}_{\rm A}$ operator}
\label{subsec: SWAP A}

In this section, we discuss the role of ${\rm SWAP}_{\rm A}$ in the spin $3/2\otimes3/2$ system. We expand the projected states $\mathcal{J}_0\ket{\psi_{\rm in}}$ and $\mathcal{J}_2\ket{\psi_{\rm in}}$ as
\begin{align}
    \mathcal{J}_0\ket{\psi_{\rm in}}
    &=
    \sum_{i,j}
    \beta_{ij} \ket{ij},
    \quad 
    \mathcal{J}_2\ket{\psi_{\rm in}}
    =
    \sum_{i,j}
    \gamma_{ij} \ket{ij},
    \label{eq: beta ij}
\end{align}
where $i,j$ run from 1 to 4, and $\beta_{ij}=\bra{ij}\mathcal{J}_0\ket{\psi_{\rm in}}$ and $\gamma_{ij}=\bra{ij}\mathcal{J}_2\ket{\psi_{\rm in}}$ are antisymmetric coefficients satisfying $\beta_{ij}=-\beta_{ji}$ and $\gamma_{ij}=-\gamma_{ji}$. From Eq.~\eqref{eq: def of in}, $\beta_{ij}$ and $\gamma_{ij}$ can be expressed in terms of $u_i$, $v_j$, and the Clebsch-Gordan (CG) coefficients.
For instance, if the third component of the total spin of the initial state $\ket{\psi_{\rm in}}$ is $J_3=0$, the projected states read
\begin{align}
    \mathcal{J}_0\ket{\psi_{\rm in},J_3=0}
    &=
    C_{0}
    \left(\frac{1}{2}\ket{14}-\frac{1}{2}\ket{23}
    +\frac{1}{2}\ket{32}-\frac{1}{2}\ket{41}
    \right),\\
    \mathcal{J}_2\ket{\psi_{\rm in},J_3=0}
    &=
    C_{2}
    \left(\frac{1}{2}\ket{14}+\frac{1}{2}\ket{23}
    -\frac{1}{2}\ket{32}-\frac{1}{2}\ket{41}\right), \\
    C_{0}
    &=\frac{1}{2}(u_{1}v_{4}-u_{2}v_{3}+u_{3}v_{2}-u_{4}v_{1}), \\
    C_{2}
    &=\frac{1}{2}(u_{1}v_{4}+u_{2}v_{3}-u_{3}v_{2}-u_{4}v_{1}),
\end{align}
and we find the following relations:
\begin{align}
    \beta_{14} = \beta_{32}
    &=
    -\beta_{41} = -\beta_{23}
    =\frac{C_{0}}{2},
    \quad
    \gamma_{14} = \gamma_{23}
    =
    -\gamma_{32} = -\gamma_{41}
    =\frac{C_{2}}{2}.
    \label{eq: rel 1}
\end{align}
These relations are specific to the spin $3/2\otimes3/2$ system and do not hold in other systems that possess both $J=0$ and $J=2$ components, such as the spin $1\otimes1$ system in $dd$ scattering.

Note that the coefficients $\beta_{ij}$ vanish except for those listed in Eq.~\eqref{eq: rel 1}, because the operator $\mathcal{J}_0$ projects onto the $J_{3}=0$ state. In other words, for the states $\ket{\psi_{\rm in},J_3\neq 0}$, the action of the ${\rm SWAP}_{\rm A}$ operator reduces to a simple antisymmetrization described by $\mathcal{A}$. For the states with $J_{3}=0$, the action of the operator $\mathcal{A}=\mathcal{J}_0+\mathcal{J}_2$ defined in Eq.~\eqref{eq: def of A} is given by
\begin{align}
    \mathcal{A}\ket{\psi_{\rm in},J_{3}=0}
    &=
    \sum_{i,j}(\gamma_{ij}+\beta_{ij})\ket{ij}
    \label{eq: gplusbeta} \\
    &=
    (\gamma_{14}+\beta_{14})\ket{14}
    + (\gamma_{23}+\beta_{23})\ket{23}
    \notag \\
    &\quad 
    +
    (\gamma_{32}+\beta_{32})\ket{32}
    + (\gamma_{41}+\beta_{41})\ket{41}.
    \label{eq: ex A}
\end{align}
On the other hand, the action of the operator ${\rm SWAP}_{\rm A}=\mathcal{J}_2-\mathcal{J}_0$ can be written as
\begin{align}
    {\rm SWAP}_{\rm A}\ket{\psi_{\rm in},J_{3}=0}
    &=
    \sum_{i,j}(\gamma_{ij}-\beta_{ij})\ket{ij}
    \label{eq: gminusbeta} \\
    &=
    (\gamma_{14}-\beta_{14})\ket{14}
    + (\gamma_{23}-\beta_{23})\ket{23}
    \notag \\
    &\quad 
    +
    (\gamma_{32}-\beta_{32})\ket{32}
    + (\gamma_{41}-\beta_{41})\ket{41}.
    \label{eq: ex SWAPA}
\end{align}
Using the relations in Eq.~\eqref{eq: rel 1}, ${\rm SWAP}_{\rm A}\ket{\psi_{\rm in},J_{3}=0}$ can be rewritten as
\begin{align}
    {\rm SWAP}_{\rm A}\ket{\psi_{\rm in},J_{3}=0}
    &=
    (\gamma_{23}+\beta_{23})\ket{14}
    + (\gamma_{14}+\beta_{14})\ket{23}
    \notag \\
    &\quad 
    +
    (\gamma_{41}+\beta_{41})\ket{32}
    + (\gamma_{32}+\beta_{32})\ket{41}.
    \label{eq: re ex A}
\end{align}
By comparing Eqs.~\eqref{eq: ex A} and~\eqref{eq: re ex A}, we find that the operator ${\rm SWAP}_{\rm A}$ effectively exchanges the single-particle basis states as
\begin{align}
    \ket{1} \leftrightarrow \ket{2}, \qquad
    \ket{3} \leftrightarrow \ket{4},
\end{align}
within the antisymmetric $J_3=0$ subspace of the spin $3/2\otimes3/2$ system.

\section{Linear entropy and $|\delta_{02}|=\pi/2$ solution}
\label{eq: sp solu}

In this section, we discuss how the minimal EP solution $|\delta_{02}|=\pi/2$ arises in $\Omega\Omega$ scattering, based on an evaluation of the linear entropy. First, in Sec.~\ref{subsec:general}, we present the general expression for the linear entropy of a state obtained by an $S$-matrix with two spin components and examine its dependence on the phase-shift difference. Next, in Sec.~\ref{subsec:EEA}, we show that the linear entropy of the properly normalized antisymmetric state is a constant, independent of the initial state. Finally, in Sec.~\ref{eq:EPOmegaOmega}, we demonstrate that the special solution $|\delta_{02}|=\pi/2$ for the minimal EP is allowed in $\Omega\Omega$ scattering due to the special properties of the CG coefficients. We also show that the minimal values of the EP for $|\delta_{02}|=0$ and $\pi/2$ can be directly inferred from the evaluation of the linear entropy.

\subsection{Linear entropy for two-component scattering}
\label{subsec:general}

In Sec.~\ref{sec:ESOmegaOmega}, we have seen that the EP for $dd$ scattering contains a $\cos[2\delta_{02}^{dd}]$ term, while such a term is absent in the EP for $\Omega\Omega$ scattering. To understand this difference, we consider the general form of the linear entropy for systems with two spin channels labeled by $J$ and $J^{\prime}$. The $S$-matrix can be written as
\begin{align}
    \hat{S}^{\rm G}
    =
    e^{2i\delta_{J}}\mathcal{J}_J
    +
    e^{2i\delta_{J^{\prime}}}\mathcal{J}_{J^{\prime}},
    \label{eq: Smat gen}
\end{align}
where $\mathcal{J}_J$ and $\mathcal{J}_{J^{\prime}}$ are the projection operators, and $\delta_J$ and $\delta_{J^{\prime}}$ are the corresponding phase shifts. For $\Omega\Omega$ and $dd$ scatterings, these correspond to $J=0$ and $J^{\prime}=2$. This $S$-matrrix also applies to $NN$ scattering with $J=0$ and $J^{\prime}=1$. The linear entropy of the normalized state defined in Eqs.~\eqref{eq: norm final} and~\eqref{eq:EPOmega} for this $S$-matrix can be expressed as
\begin{align}
    \mathcal{E}(\hat{S}^{\rm G}\ket{\psi_{\rm in}}/N)
    &=
    \mathcal{E}((\mathcal{J}_J+\mathcal{J}_{J^{\prime}})\ket{\psi_{\rm in}}/N)
    \notag \\
    &\quad 
    - \frac{1}{N^4}\left\{
    2(\cos[2\delta_{JJ^{\prime}}]-1)\Tr_1[(\rho^{J}_1+\rho^{J^{\prime}}_1)(\rho^{\prime}_1+\rho^{\prime\dagger}_1)]
    \right.
    \notag \\
    &\quad +2i\sin[2\delta_{JJ^{\prime}}]\Tr_1[(\rho^{J}_1+\rho^{J^{\prime}}_1)(\rho^{\prime}_1-\rho^{\prime\dagger}_1)]
    \notag \\
    &\quad 
    +
    (\cos[4\delta_{JJ^{\prime}}]-1)\Tr_1[(\rho^{\prime}_1)^2+(\rho^{\prime\dagger}_1)^2]
    \notag \\
    &\quad \left. +
    i\sin[4\delta_{JJ^{\prime}}]\Tr_1[(\rho^{\prime}_1)^2-(\rho^{\prime\dagger}_1)^2]\right\}
    ,
    \label{eq: LE for Omega}
\end{align}
where the normalization constant $N$, the phase-shift difference $\delta_{JJ^{\prime}}$, and the density operators are defined as
\begin{align}
    N&=\sqrt{\bra{\psi_{\rm in}}(\mathcal{J}_J+\mathcal{J}_{J^{\prime}})\ket{\psi_{\rm in}}},
    \quad
    \delta_{JJ^{\prime}}
    =
    \delta_{J} - \delta_{J^{\prime}}, \\
    \rho^{J}_{1} &\equiv 
    \Tr_{2}
    \left[\mathcal{J}_J\ket{\psi_{\rm in}}\bra{\psi_{\rm in}}\mathcal{J}_J\right],
    \quad 
    \rho^{\prime}_{1} \equiv 
    \Tr_{2}
    \left[\mathcal{J}_J\ket{\psi_{\rm in}}\bra{\psi_{\rm in}}\mathcal{J}_{J^{\prime}}\right].
\end{align}

The expression of the normalized entropy in Eq.~\eqref{eq: LE for Omega} contains trigonometric functions with arguments $2\delta_{JJ^{\prime}}$ and $4\delta_{JJ^{\prime}}$. This indicates that the EP calculated from the entropy~\eqref{eq: LE for Omega} can depend on the phase-shift difference $\delta_{JJ^{\prime}}$ only through these combinations. The terms depending on $2\delta_{JJ^{\prime}}$ in the entropy produce a $\cos[2\delta_{JJ^{\prime}}]$ term in the EP after averaging over the initial states, whereas those depending on $4\delta_{JJ^{\prime}}$ give rise to the $\cos[4\delta_{JJ^{\prime}}]$ term in the EP. This behavior is explicitly confirmed in the EP for $dd$ scattering in Eq.~\eqref{eq: EF for dd}, which exhibits such a dependence on $\delta_{JJ^{\prime}}$. Moreover, the absence of the $\cos[2\delta_{JJ^{\prime}}]$ term in the EP for $\Omega\Omega$ scattering~\eqref{eq: EP for omega}, which leads to the $|\delta_{02}|=\pi/2$ solution and the ${\rm SWAP}_{\rm A}$ structure, implies the absence of the $2\delta_{JJ^{\prime}}$ terms in the entropy. In the following, we examine how these terms vanish in $\Omega\Omega$ scattering.

\subsection{Linear entropy for antisymmetric state}
\label{subsec:EEA}

To evaluate the linear entropy for $\Omega\Omega$ scattering, we first focus on the first term in Eq.~\eqref{eq: LE for Omega}. For $\Omega\Omega$ scattering, the sum of the relevant projection operators is the antisymmetrization operator,
$\mathcal{J}_J+\mathcal{J}_{J^{\prime}}=\mathcal{J}_0+\mathcal{J}_{2}=\mathcal{A}$.
Therefore, the first term in Eq.~\eqref{eq: LE for Omega} corresponds to the linear entropy of the normalized antisymmetric state,
\begin{align}
    \frac{(\mathcal{J}_J+\mathcal{J}_{J^{\prime}})\ket{\psi_{\rm in}}}
    {N}
    =
    \frac{\mathcal{A}\ket{\psi_{\rm in}}}
    {\sqrt{\bra{\psi_{\rm in}}\mathcal{A}\ket{\psi_{\rm in}}}}
    \equiv \ket{\psi^{\rm A}} .
    \label{eq:psiA}
\end{align}
Using the product-state expression of $\ket{\psi_{\rm in}}$ in Eq.~\eqref{eq: def of in}, the normalized antisymmetric state can be written in terms of $u_i$ and $v_j$ as
\begin{align}
    \ket{\psi^{\rm A}}
    &=
    \sum_{i,j}
    \alpha_{ij}^{\rm A}
    \ket{ij},
    \quad
    \alpha_{ij}^{\rm A}
    =
    \frac{u_iv_j-u_jv_i}
    {2\sqrt{\bra{\psi_{\rm in}}\mathcal{A}\ket{\psi_{\rm in}}}} .
    \label{eq: alphaA} 
\end{align}

To evaluate the linear entropy~\eqref{eq: entropy general} of $\ket{\psi^{\rm A}}$, we first note that
\begin{align}
    \qty|
    \alpha_{ij}^{\rm A}\alpha_{i^{\prime}j^{\prime}}^{\rm A}
    -
    \alpha_{i^{\prime}j}^{\rm A}\alpha_{ij^{\prime}}^{\rm A}
    |
    &=
    \qty|u_iv_{i^{\prime}}-u_{i^{\prime}}v_i|^2 .
\end{align}
Using this relation, the linear entropy of $\ket{\psi^{\rm A}}$ is evaluated as
\begin{align}
    \mathcal{E}(\ket{\psi^{\rm A}})
    &=
    \frac{1}{2\bra{\psi_{\rm in}}\mathcal{A}\ket{\psi_{\rm in}}^2}
    \qty(
        \sum_{i<i^{\prime}}
        \frac{\qty|u_iv_{i^{\prime}}-u_{i^{\prime}}v_i|^2}{2}
    )^2
    =\frac{1}{2}.
    \label{eq: nibuichi}
\end{align}
Here we have used Eq.~\eqref{eq: norm A}. The resulting linear entropy takes the constant value $1/2$, independent of the choice of the initial state $\ket{\psi_{\rm in}}$. We further note that this value coincides with the minimal EP for $\Omega\Omega$ scattering, realized at $|\delta_{02}|=0$ and $\pi/2$. In the next subsection, we demonstrate that Eq.~\eqref{eq: nibuichi} indeed gives the minimal value of EP.

\subsection{Linear entropy and entanglement power for $\Omega\Omega$ scattering}
\label{eq:EPOmegaOmega}

We are now in a position to evaluate the linear entropy for $\Omega\Omega$ scattering using Eq.~\eqref{eq: LE for Omega}. We first show that the terms proportional to $2\delta_{02}$ vanish in the linear entropy for $\Omega\Omega$ scattering. This originates from the special relation~\eqref{eq: rel 1} among the Clebsch-Gordan coefficients. In Eq.~\eqref{eq: LE for Omega}, the $2\delta_{02}$ terms are proportional to $\Tr_1[(\rho^{J=0}_1+\rho^{J=2}_1)(\rho^{\prime}_1\pm\rho^{\prime\dagger}_1)]$, which can be expressed in terms of the coefficients $\beta_{ij}$ and $\gamma_{ij}$ defined in Eq.~\eqref{eq: beta ij} as
\begin{align}
    \Tr_1[(\rho^{J=0}_1+\rho^{J=2}_1)(\rho^{\prime}_1\pm\rho^{\prime\dagger}_1)]
    &=
    \sum_{i,j,i^{\prime},j^{\prime}}
    \qty(\beta_{ij}\beta_{i^{\prime}j}^*+\gamma_{ij}\gamma_{i^{\prime}j}^*)
    \qty(\beta_{i^{\prime}j^{\prime}}\gamma_{ij^{\prime}}^*\pm\beta_{ij^{\prime}}^*\gamma_{i^{\prime}j^{\prime}})
    \notag \\
    &=
    \Bigl[
    (|\gamma_{12}|^2+|\gamma_{13}|^2+|\gamma_{24}|^2+|\gamma_{34}|^2)
    \notag \\
    &\quad
    -
    (|\gamma_{12}|^2+|\gamma_{24}|^2+|\gamma_{13}|^2+|\gamma_{34}|^2)
    \Bigr]
    \notag \\
    &\quad \times
    \qty(\beta_{14}\gamma_{14}^*+\beta_{14}^*\gamma_{14})
    \notag \\
    &=
    0,
    \label{eq: zero con}
\end{align}
where we have used the relation~\eqref{eq: rel 1}. Therefore, the absence of the $2\delta_{02}$ dependence in the entanglement power for $\Omega\Omega$ scattering originates from the special structure of the CG coefficients. This conclusion is consistent with the discussion in Sec.~\ref{subsec: SWAP A}, where the relation~\eqref{eq: rel 1} plays a crucial role in the interpretation of the ${\rm SWAP}_{\rm A}$ operator.

From Eqs.~\eqref{eq: LE for Omega}, \eqref{eq:psiA}, and~\eqref{eq: zero con}, the linear entropy for $\Omega\Omega$ scattering can now be evaluated as
\begin{align}
    \mathcal{E}(\ket{\psi^{\Omega\Omega}_{\rm out}})
    &=
    \mathcal{E}(\ket{\psi^{\rm A}})
    \notag \\
    &\quad 
    - \frac{1}{\bra{\psi_{\rm in}}\mathcal{A}\ket{\psi_{\rm in}}^2}\left\{
    \qty(\cos[4\delta_{02}]-1)\Tr_1[(\rho^{\prime}_1)^2+(\rho^{\prime\dagger}_1)^2]
    \right.
    \notag \\
    &\quad \left. +
    i\sin[4\delta_{02}]\Tr_1[(\rho^{\prime}_1)^2-(\rho^{\prime\dagger}_1)^2]\right\},
    \label{eq: LE for Omega correct} 
\end{align}
which exhibits only a $4\delta_{02}$ dependence. 
Note that $\bra{\psi_{\rm in}}\mathcal{A}\ket{\psi_{\rm in}}$, $\Tr_1[(\rho^{\prime}_1)^2]$, and $\Tr_1[(\rho^{\prime\dagger}_1)^2]$ depend on the spin directions of the initial state $\ket{\psi_{\rm in}}$. Therefore, to obtain an explicit expression for the EP, one must perform the angular integration over the initial spin directions. Nevertheless, Eq.~\eqref{eq: LE for Omega correct} already demonstrates that the dependence of the EP on the phase shift difference is constrained to $4\delta_{02}$. In other words, terms with $2\delta_{02}$ dependence are forbidden in the EP for $\Omega\Omega$ scattering.

Recall that the minimal EP solutions are given by $|\delta_{02}|=0,\pi/2$. In these cases, since $\cos[4\delta_{02}]=1$ and $\sin[4\delta_{02}]=0$, the linear entropy $\mathcal{E}(\ket{\psi^{\Omega\Omega}_{\rm out}})$ becomes independent of the initial spin directions and is evaluated as
\begin{align}
    \mathcal{E}(\ket{\psi^{\Omega\Omega}_{\rm out}})
    &=
    \frac{1}{2},
    \quad
    (|\delta_{02}|=0,\pi/2),
    \label{eq: min sol}
\end{align}
thanks to Eq.~\eqref{eq: nibuichi}. This result shows that, for $|\delta_{02}|=0,\pi/2$, the linear entropy always takes the value $1/2$, independently of the initial state $\ket{\psi_{\rm in}}$. Accordingly, the EP, obtained by averaging the linear entropy over the initial state, is also equal to $1/2$. In this manner, the EP for the cases $|\delta_{02}|=0,\pi/2$ can be directly determined from the constant linear entropy in Eq.~\eqref{eq: min sol}. Moreover, since the integrand of the normalized EP defined in Eq.~\eqref{eq:EPOmega} is independent of the angles $\omega_{1}$ and $\omega_{2}$, the quantity $E_{2}$ defined in Eq.~\eqref{eq:EP2Omega}, as well as any $k$-th weighted EPs introduced in Ref.~\citen{Hu:2025lua}, always yields $1/2$ for $|\delta_{02}|=0,\pi/2$.

As mentioned at the beginning of Sec.~\ref{subsec:general}, the expression of the linear entropy in Eq.~\eqref{eq: LE for Omega} is also valid for $NN$ scattering. In this case, the first term vanishes because $\mathcal{J}_J+\mathcal{J}_{J^{\prime}}=\mathcal{J}_0+\mathcal{J}_{1}=\hat{1}$. Furthermore, the $2\delta_{JJ^{\prime}}$ terms vanish, since they are proportional to the trace of the product of an odd number of antisymmetric matrices. As a result, the minimal EP $E(\hat{S})=0$ is obtained for 
$|\delta_{01}|=0,\pi/2$.

\section{Summary}

We have investigated entanglement suppression in hadronic scattering by focusing on $\Omega\Omega$ scattering, a system of identical particles with spin $3/2$. First, in Sec.~\ref{sec: formulation}, we review entanglement suppression as a guiding principle for identifying emergent symmetries in $NN$ scattering. It is shown that the identity and SWAP gates emerge as the $S$-matrix solutions that minimize the EP in $NN$ scattering, leading to the spin-flavor SU(4) symmetry and nonrelativistic conformal symmetry, respectively.

The entanglement suppression for $\Omega\Omega$ scattering is discussed in Sec.~\ref{sec: Omega application}. In this system, due to the antisymmetrization of identical fermions, the spin-$3/2$ baryons scatter through the spin $J=0$ and $J=2$ channels. Minimization of the EP admits two solutions, $|\delta_{02}|=0$ and $|\delta_{02}|=\pi/2$. The solution $|\delta_{02}|=0$ leads to an $S$-matrix proportional to the identity operator within the antisymmetric subspace, which is associated with the spin SU(4) symmetry. On the other hand, the solution $|\delta_{02}|=\pi/2$ is realized by the ${\rm SWAP}_{\rm A}$ $S$-matrix and is associated with the nonrelativistic conformal symmetry. We demonstrate that the ${\rm SWAP}_{\rm A}$ operator exchanges the spin components within the antisymmetric $J_3=0$ subspace.

Finally, in Sec.~\ref{eq: sp solu}, we explicitly compute the dependence of the linear entropy for $\Omega\Omega$ scattering on the phase shift difference. We find that the specific properties of the CG coefficients in the $3/2 \otimes 3/2$ system eliminate the $2\delta_{02}$ dependence, thereby allowing the minimal EP solution $|\delta_{02}|=\pi/2$. Furthermore, we show that the normalized linear entropy equals $1/2$, independently of the initial state, for $|\delta_{02}|=0$ and $\pi/2$. This demonstrates that the minimal EP in $\Omega\Omega$ scattering can be directly evaluated from the linear entropy once the condition on $\delta_{02}$ is identified.

\section*{Acknowledgments}

KS is supported in part by JSPS KAKENHI Grant Number JP25KJ1996, and by MIYAKO-MIRAI Project of Tokyo Metropolitan University;
TRH is supported in part by the China National Training Program of Innovation and Entrepreneurship for Undergraduates under Grant No.~202414430004;
TRH and FKG are supported in part by the National Natural Science Foundation of China under Grants No. 12125507, No. 12361141819, and No. 12447101; and by the Chinese Academy of Sciences under Grants No.~YSBR-101;
TH is supportedd in part by the Grants-in-Aid for Scientific Research from JSPSs (Grants No.~JP23H05439, and No.~JP22K03637);
IL is supported in part by the U.S. Department of Energy under contracts DE-AC02-06CH11357, DE-SC0023522, DE-SC0010143  and No. 89243024CSC000002 (QuantISED Program)

\section*{ORCID}


\noindent Katsuyoshi Sone - \url{https://orcid.org/0000-0002-2755-3284}

\noindent Tao-Ran Hu - \url{https://orcid.org/0009-0003-9720-0171}

\noindent Tetsuo Hyodo - \url{https://orcid.org/0000-0002-4145-9817}

\noindent Feng-Kun Guo - \url{https://orcid.org/0000-0002-2919-2064}

\noindent Ian Low - \url{https://orcid.org/0000-0002-7570-9597}


%

\end{document}